\documentclass[amsmath,amssymb,twocolumn,pre]{revtex4}
\usepackage{graphicx}
\usepackage{dcolumn}
\usepackage{subfigure}
\usepackage{bm}
\usepackage{color}
\usepackage{braket}

\begin{document}

\title{Topological transitions in the configuration space of non-Euclidean origami}
\author{M. Berry}
\author{M.E. Lee-Trimble}
\affiliation{University of Massachusetts, Amherst, MA 01003 USA}
\author{C.D. Santangelo}
\email{cdsantan@syr.edu}
\altaffiliation{now at Syracuse University, Syracuse, NY,  13244 USA}
\affiliation{University of Massachusetts, Amherst, MA 01003 USA}

\date{\today}

\begin{abstract}
Origami structures have been proposed as a means of creating three-dimensional structures from the micro- to the macroscale, and as a means of fabricating mechanical metamaterials. The design of such structures requires a deep understanding of the kinematics of origami fold patterns. Here, we study the configurations of non-Euclidean origami, folding structures with Gaussian curvature concentrated on the vertices. The kinematics of such structures depends crucially on the sign of the Gaussian curvature. The configuration space of non-intersecting, oriented vertices with positive Gaussian curvature decomposes into disconnected subspaces; there is no pathway between them without tearing the origami. In contrast, the configuration space of negative Gaussian curvature vertices remain connected. This provides a new mechanism by which the mechanics and folding of an origami structure could be controlled.
\end{abstract}

\maketitle


Origami and kirigami have been proposed as a framework to engineer new materials with complex mechanical responses \cite{fuchi2012origami, schenk2013geometry, wei2013geometric,silverberg2014using,castle2014making}. To this end, new fabrication methods have been developed to enable the folding of three dimensional structures from thin films \cite{na2015programming,callens2018,plucinsky2018}. Though most examples of origami structures are foldable from an initially flat sheet, two  threads of research suggest a need to understand the motions of a broader class of ``curved'' origami. First, kirigami structures, initially flat structures with holes which can be glued together along their free edges to create intrinsically buckled structures \cite{castle2014making}. Second, newer origami fabrication methods have enabled vertices with Gaussian curvature and curved faces \cite{plucinsky2018,garza2019,Alperin2012,bende2015}.

\begin{figure}[b]
\includegraphics[width=3.5in]{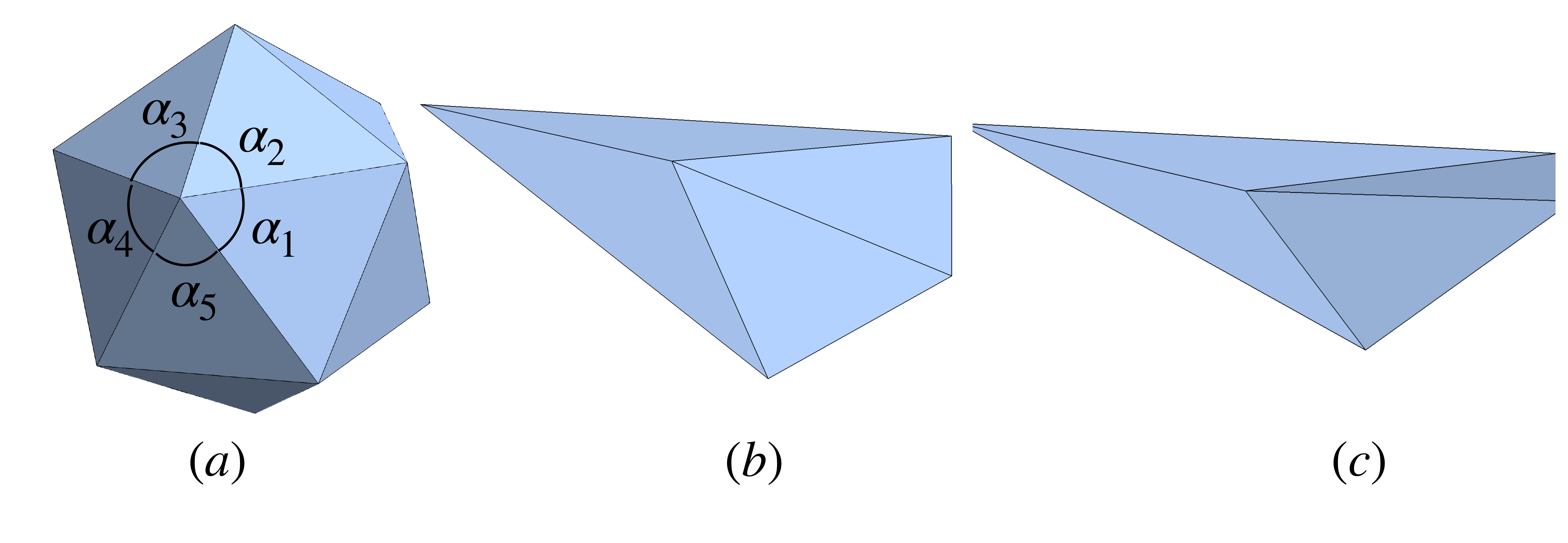}
\caption{\label{fig:what} (a) A generic non-Euclidean origami structure. The vertex Gaussian curvature is defined by $K = 2 \pi - \sum_i \alpha_i$ (b,c) Degree four vertices with positive and negative Gaussian curvatures respectively necessarily buckle out of the plane.}
\end{figure}

This paper analyzes the kinematics of non-Euclidean origami in the limit that it is almost flat. By ``non-Euclidean origami,'' we mean that faces are flat, but that the vertices have Gaussian curvature (Fig. \ref{fig:what} a--c). This Gaussian curvature manifests as either a deficit or excess angle when summing the sector angles around the internal vertices (Fig. \ref{fig:what}a). By ``almost flat,'' we mean that both the sum of sector angles around internal vertices is near $2 \pi$ and that the dihedral angles of the folds are nearly $\pi$. In this limit, we will develop a general framework for studying origami motions, and make contact with both the kinematics of flat origami structures \cite{chen2018} and continuum equations governing the small deformations of elastic sheets \cite{seung1988}.

Understanding whether an origami fold pattern can be folded without tearing is NP-hard \cite{akitaya2018rigid}. More generally, when mapping out the space of possible configurations of a given origami fold pattern, the configuration space can be geometrically complex. Additionally, these spaces can undergo topological changes as the fold pattern changes that lead to changes in the mechanical properties of origami \cite{liu2018topological}.
Here, we show that vertex Gaussian curvature can induce a topological change in the configuration space of origami structures. In particular, we will show that origami with positive Gaussian curvature vertices have configuration spaces that become disconnected, and that such disconnection need not (and likely does not) occur for negative Gaussian curvature.

\section{Mathematical Formulation}

We model origami by a collection of polygonal faces meeting at point-like vertices and joined along line-like, rigid edges, as shown in Fig. \ref{fig:what} for triangular faces. We find it useful to distinguish internal vertices, whose number we will denote $V_i$, from boundary vertices, whose number is $V_b$. Note that in traditional origami nomenclature a ``vertex'' denotes only the internal vertices. Similarly, we denote the internal and boundary edges by $E_i$ and $E_b$, respectively. The internal edges are the folds in the origami literature.

We are primarily interested in determining the isometries of a given origami fold pattern, \textit{i.e.} the motions that preserve the length of all edges and the angles between any two adjacent edges on the same face. In the case of triangular faces, the angle constraint is redundant -- once the length of all the edges are known, the angles between edges are already uniquely determined. Thus, we will focus mainly on origami with triangular faces. This is not very restrictive; we will see that the configuration space of an origami structure with polygonal faces can be obtained by taking a lower dimensional slice through the configuration space of a suitable triangulated origami fold pattern. To define the discrete Gaussian curvature of an internal vertex, we measure the sector angles, $\alpha_{i}$, between adjacent folds with one end on a given vertex (Fig. \ref{fig:what}a). The Gaussian curvature of that vertex is then $K_n = 2\pi-\sum_i \alpha_{i}$ \cite{meyer2003discrete}.

\begin{figure}[b]
\includegraphics[width=3.5in]{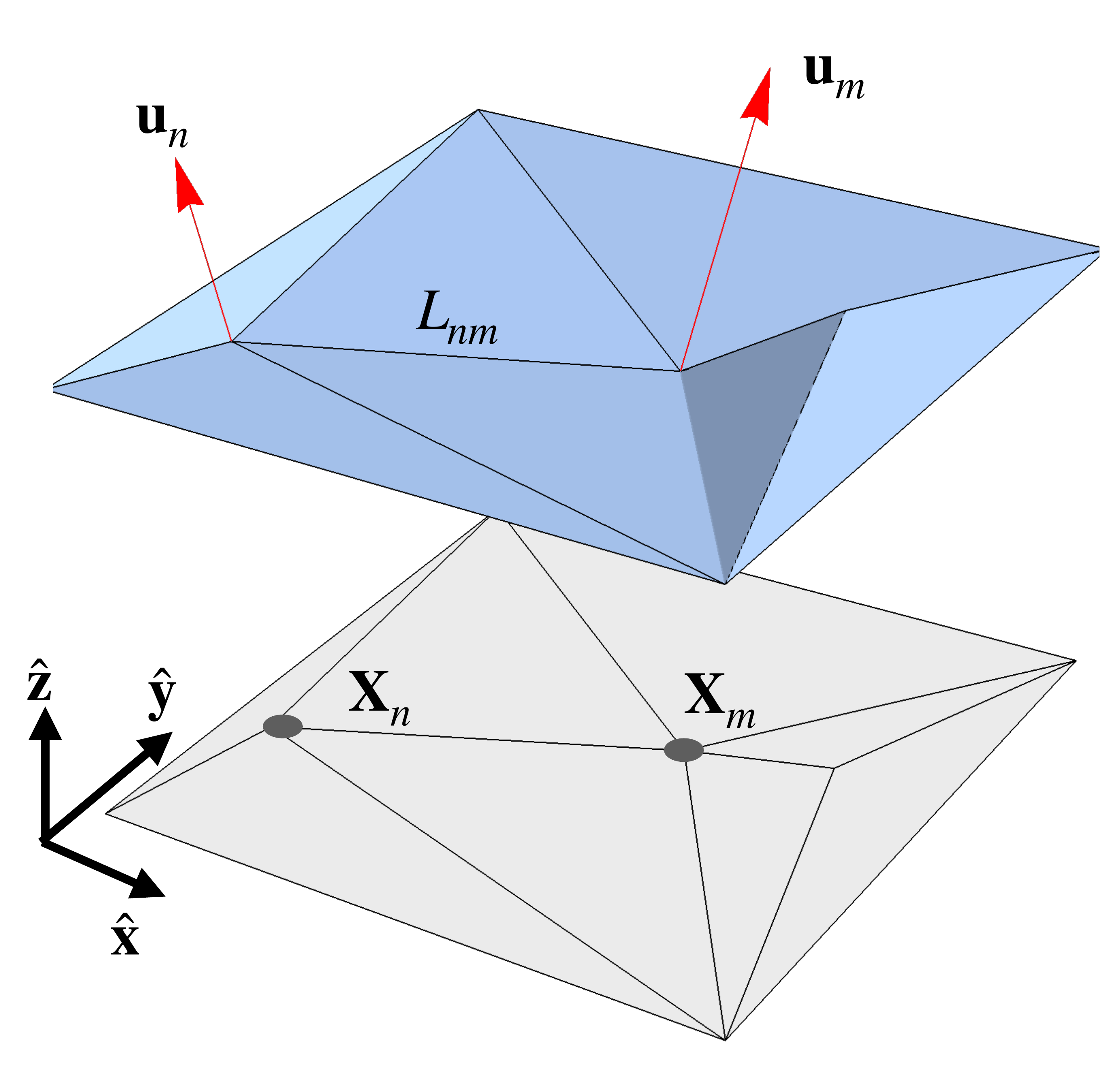}
\caption{\label{fig:notation} A nearly flat origami structure can be projected to a fold pattern in the $xy-$plane. In-plane and out-of-plane displacements are unambiguously decomposable.}
\end{figure}

One of the primary features of triangulated origami is that the number of infinitesimal isometries is almost precisely balanced by the number of constraints. This is true for any Gaussian curvature though it manifests in different ways when $K_n = 0$ on each internal vertex. Understanding this distinction turns out to be important to developing a fuller picture of the origami configuration space so we review it here. If $\mathbf{X}_n$ denotes the three dimensional position of the $n^{th}$ vertex, then any pair of vertices joined by an edge induces a geometrical constraint,
\begin{equation}\label{eq:fulleq}
(\mathbf{X}_n - \mathbf{X}_m)^2 = L_{nm}^2,
\end{equation}
where $L_{nm}$ is the length of the edge between $n$ and $m$. We then write $\mathbf{u}_n$ (Fig. \ref{fig:notation}b) as the displacement of the $n^{th}$ vertex, and find that, to first order, motions are governed by the linear equations
\begin{equation}\label{eq:linearisom}
2 (\mathbf{X}_n - \mathbf{X}_m) \cdot (\mathbf{u}_n-\mathbf{u}_m) = 0.
\end{equation}
There is one equation of this type for each edge $(n,m)$ joining vertex $n$ to $m$.

To understand the generic behavior of Eq. (\ref{eq:linearisom}), we note that there are $E_i + E_b$ constraints, one for each edge and $3 V_i + 3 V_b$ naive degrees of freedom associated with the three-dimensional displacements of the vertices. A triangulated origami fold pattern also satisfies both Euler's theorem, $F-E_i - E_b + V_i + V_b = 1$, where $F$ is the number of faces, and satisfies the $2 E_i+E_b = 3 F$ to account for the fact that each face is associated to three edges but each internal edge joins two faces. Similarly, we have $E_b = V_b$ because the boundary of the fold pattern is a polygon. Taken together, these equations imply $E_i = V_b + 3 V_i - 3$ and so naive counting suggests that the dimension of the configuration space of origami is $D = V_b + 3$. Six of these degrees of freedom are Euclidean motions.

Though this generic counting should be valid for most configurations, it fails when the origami is flat because the constraints at first order are not all independent. In that case, only the in-plane deformations are fixed by the length constraints: any vertex can be displaced vertically without causing a first-order change in the edge lengths. Though this suggests that $D = V_i + V_b + 3$, it turns out that there are additional constraints at quadratic order in the lengths, If we define $\mathbf{h} = (h_1,h_2,\cdots)$ as a vector specifying the vertical displacement of each of the vertices above the $xy-$plane, then a necessary and sufficient condition for a motion to be an isometry to second order is
\begin{equation}\label{eq:noK}
\mathbf{h}^T \mathbf{Q}_n \mathbf{h} = 0,
\end{equation}
for each internal vertex, $n$, where the matrix $\mathbf{Q}_n$ depends on the sector angles of internal vertex $n$ \cite{chen2018}. The left-hand side of Eq. (\ref{eq:noK}) is the Gaussian curvature of internal vertex $n$ induced by the height changes \cite{chen2018} so Eq. (\ref{eq:noK}) is simply the statement that no infinitesimal deformation can change the Gaussian curvature of the internal vertices. There are precisely enough quadratic constraints, one for each internal vertex, to recover the generic result, $D= V_b + 3$.


We now wish to modify Eq. (\ref{eq:noK}) to allow for internal vertices to have a small but nonzero Gaussian curvature. In this regime, the vertices continue to remain almost planar and, in Appendix \ref{sec:formalism}, we show that the geometrical constraints at each vertex should be modified to
\begin{equation}\label{eq:mainresult}
\mathbf{h}^T \mathbf{Q}_n \mathbf{h} = K_n,
\end{equation}
where $K_n$ is the Gaussian curvature of vertex $n$ and $\mathbf{Q}_n$ is the same matrix that appears in Eq. (\ref{eq:noK}) for flat origami. Despite the plausible form of Eq. (\ref{eq:mainresult}), the proof that Eq. (\ref{eq:mainresult}) correctly determines the isometries to quadratic order is somewhat involved.

Eq. (\ref{eq:mainresult}) can be contrasted to the equations governing the small isometries of a continuum elastic sheet, which are governed by the approximate equations \cite{seung1988}
\begin{equation}\label{eq:gkequation}
K = -\frac{1}{2} \sum_{i j k l = 1}^2 \epsilon_{i k} \epsilon_{j l} \partial_i \partial_j h \partial_k \partial_l h,
\end{equation}
where $\epsilon_{i j}$ is the antisymmetric Levi-Civita symbol with $\epsilon_{12}=1$, $h(x,y)$ is the vertical height of the elastic sheet above the $xy-$plane, and $K(x,y)$ is the Gaussian curvature. Eq. (\ref{eq:gkequation}) is accurate in the limit of small slopes $|\partial_i h| \ll 1$, which is precisely the same limit of our discrete formulation. In that sense, Eq. (\ref{eq:mainresult}) is a discrete analogue to the better known continuum result of Eq. (\ref{eq:gkequation}).

\begin{figure}
\includegraphics[width=3in]{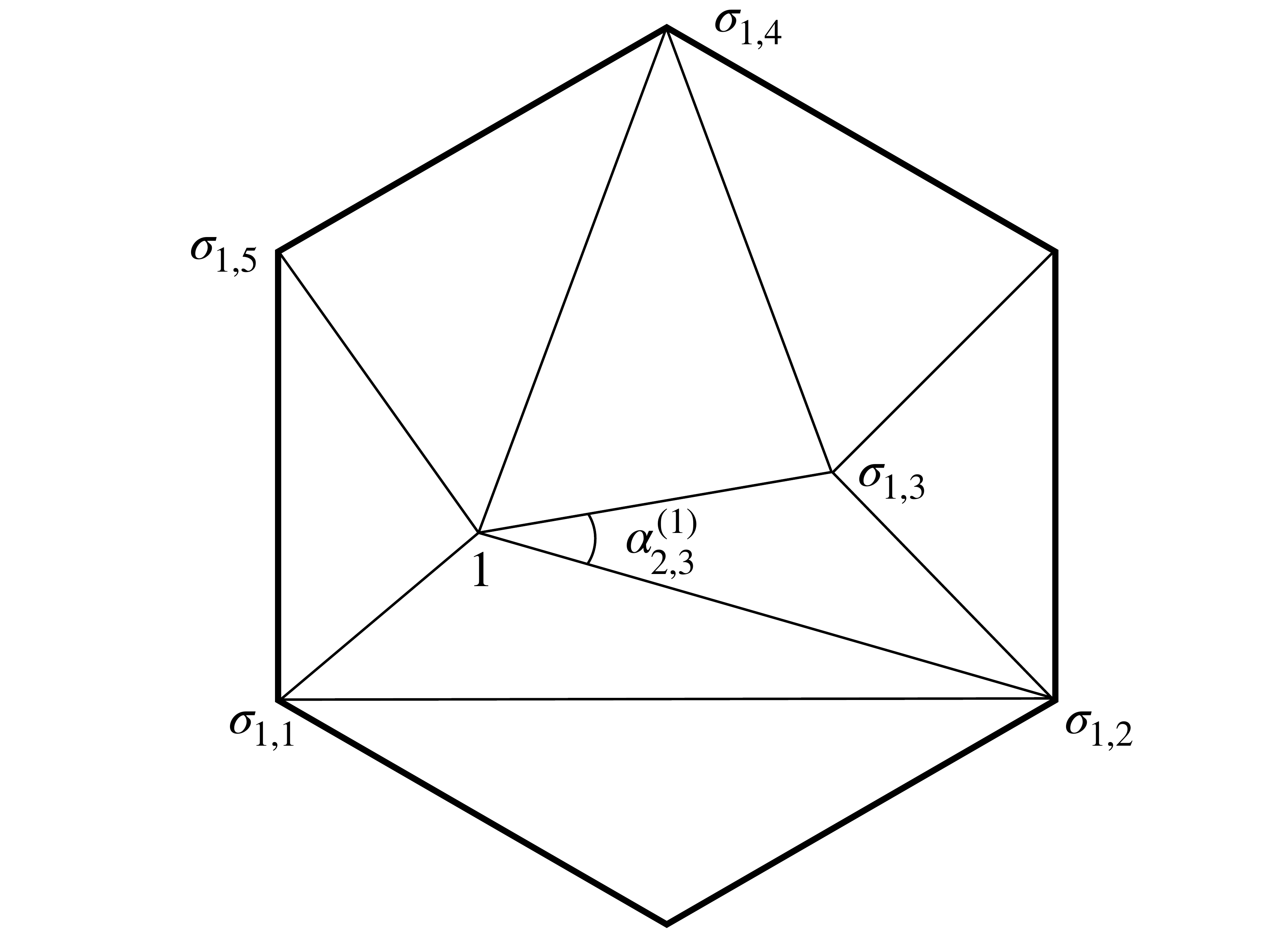}
\caption{\label{fig:notation2} Notation for the vicinity of a single vertex.}
\end{figure}

In Appendix \ref{sec:vertices}, we also show that we can rewrite the Gaussian curvature around any internal vertex $n$ in closed form. To do this consistently requires some additional notation (Fig. \ref{fig:notation2}).
Let $\sigma(n,1)$ through $\sigma(n,N(n))$ be the vertices connected to an internal vertex $n$ in counterclockwise order, where $N(n)$ is the number of edges with $n$ at one end. We also denote $L_{nm}$ as the length of the edge joining vertex $n$ to $m$. Then,
\begin{equation}\label{eq:mainresult2}
K_n = -\frac{1}{2} {\sum_i^{N(n)}}\sum_j^{N(n)} \left(\frac{h_{\sigma(n,i)}-h_n}{L_{\sigma(n,i) n}}\right) M_{i j}^{(n)} \left(\frac{h_{\sigma(n,j)}-h_n}{L_{\sigma(n,j) n}}\right).
\end{equation}
The matrix $M_{i j}^{(n)}$ is an $N(n) \times N(n)$ square matrix depending on the sector angles around each internal vertex. To define $M_{ij}^{(n)}$, let $\alpha^{(n)}_{i,i+1}$ be the sector angle between vertex $\sigma(n,i)$ and $\sigma(n,i+1)$ around the vertex $n$. Then
\begin{eqnarray}
M_{ij}^{(n)} &=& -\csc \alpha_{i,i+1}^{(n)} \delta_{i,j+1} - \csc \alpha_{i-1,i}^{(n)} \delta_{i,j-1} \\
& &+ (\cot \alpha_{i,i+1}^{(n)} + \cot \alpha_{i-1,i})^{(n)} \delta_{ij}.\nonumber
\end{eqnarray}
For sector angles smaller than $\pi$, the matrices $\mathbf{M}^{(n)}$ have two zero eigenvalues, one negative eigenvalue, and the remaining eigenvalues are positive (see Ref. \cite{kapovich1997hodge} or Appendix C of Ref. \cite{chen2018} for a detailed proof).


\section{Single vertices}

\begin{figure}
\includegraphics[width=3.5in]{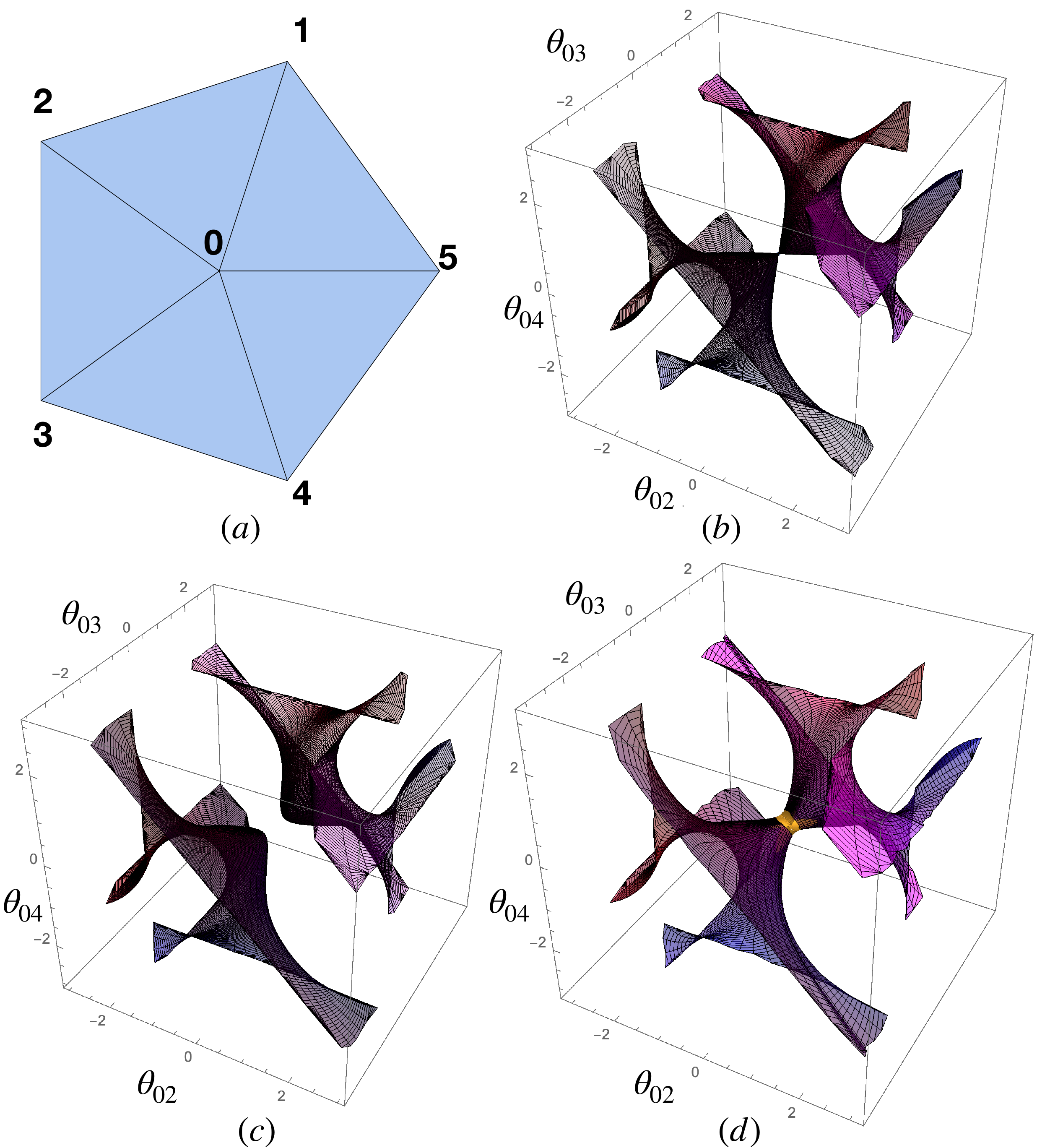}
\caption{\label{fig:config} The configuration space of a symmetric five-fold vertex (a) near the flat state with zero (b), positive (c), and negative (d) Gaussian curvature projected onto the fold angles $(\theta_{02},\theta_{03},\theta_{04})$. The fraction of red and blue, $[r,b]$, in the coloring is determined by $[(\theta_{01}-\pi)/(2 \pi),(\theta_{05}+\pi)/(2 \pi) ]$.}
\end{figure}

To better understand Eq. (\ref{eq:mainresult2}), consider an origami structure with one internal vertex from which $N$ folds emerge (Fig. \ref{fig:config}a). This case has been analyzed in some depth due to the correspondence between origami vertices of degree $N$ and spherical linkages with $N$ segments \cite{kapovich1997hodge,streinu2004}.
We denote the height of the central vertex $h_0$ and the heights of the surrounding vertices $h_1$ through $h_N$, and we explicitly eliminate rigid body motions by fixing the heights of vertices $(h_0,h_1,h_2) = (0,0,0)$. One quadratic constraint remains on the remaining heights, $h_3$ to $h_N$, leaving $N-3$ distinct degrees of freedom.

So what does the configuration space of a single non-Euclidean vertex look like? We suppose $\mathbf{e}_{-,i}$ are the components of the normalized eigenvector corresponding to the negative eigenvalue, $-\lambda_-$, of $\mathbf{M}^{(0)}$. We then suppose $\mathbf{e}_{n,i}$ are the components of the $n^{th}$ normalized eigenvector with positive eigenvalues, $\lambda_{n}$. We can then attempt to solve Eq. (\ref{eq:mainresult2}) with the ansatz
\begin{equation}
\frac{h_i - h_0}{L_{i 0}} = \frac{1}{\sqrt{2 \lambda_-}} c_- \mathbf{e}_{-,i} + \sum_n \frac{1}{\sqrt{2 \lambda_{n}}} c_n \mathbf{e}_{n,i}.
\end{equation}
We find that $c_-^2 - \sum_n c_n^2 = K_0$, where $K_0$ is the Gaussian curvature of vertex $0$.

When $K_0=0$, we recover the results of Ref. \cite{chen2018}: the solution forms a cone described by the equation $c_- = \pm \sqrt{c_n^2}$ with a singularity at $c_-=c_n=0$. Each nappe of the solution space is characterized by the sign of $c_-$ (called branch signs in \cite{chen2018}).
When $K_0 > 0$, we must instead solve,
\begin{equation}
c_-^2 = K_0 + \sum_n c_n^2,
\end{equation}
showing that $|c_-| \ge K_0$. 
This would seem to imply that the two nappes have split into two disconnected components characterized by the sign of $c_-$.
Finally, we turn to $K_0<0$, for which
\begin{equation}
c_-^2 + |K| = \sum_n c_n^2. 
\end{equation}
Here, it is clear that there is no obstruction to $c_- = 0$. Instead, $\sum_n c_n^2 \ge |K|$. We conclude that the conical configuration space is one in which both nappes remain connected near the flat state but are connected by a neck (Fig. \ref{fig:config}d). This is quite different than what happens for degree four vertices \cite{waitukaitis2016}.

In Fig. \ref{fig:config}, we numerically plot the configuration space of a symmetric degree-5 vertex. To do this, we compute radial trajectories from a known configuration of the origami vertex. Each point of the radial trajectory is found in a sequence of steps. For each step, we solve Eq. (\ref{eq:linearisom}) to identify the infinitesimal isometries from any configuration that is not flat and project the previous tangent direction onto the new tangent space. After finding a new configuration using the linear isometry, we numerically minimize the energy functional,
\begin{equation}\label{eq:energyNumerics}
E = \frac{1}{2} \sum_{nm} \left[ \left(\mathbf{X}_n - \mathbf{X}_m\right)^2-L_{nm}^2\right]^2,
\end{equation}
where the sum is over edges joining vertex $n$ to $m$, using the BFGS (``QuasiNewton'') algorithm in Mathematica 11. This prevents numerical errors in the linear isometries from building up as the integration proceeds. This process proceeds until one of the fold angles exceeds $\pi$ or $-\pi$, indicating that a face has come into contact with an adjacent face. Finally, the trajectories are assembled into a mesh to produce a surface.

Generically, we find that the configuration space near the flat state follows the analytical results we obtained. Specifically, it appears that the configuration space decomposes into two nappes with the topology of a disk which are either touching at one point ($K=0$), disconnected ($K > 0$), or connected by a narrow neck ($K<0$). At first glance, this appears to contradict Streinu and Whitely \cite{streinu2004},  who showed that the configuration space of single vertices with $K_0>0$ is always connected. However, in their analysis, faces can pass through each other; whereas in Fig. \ref{fig:config}, fold angles must remain strictly between $-\pi$ and $\pi$. Anecdotally, it does appear that when faces are allowed to pass through each other, isometric trajectories can pass from one nappe to the other for any $K$. In this case, however, the surfaces become difficult to plot, even more difficult to understand, and, in any case, are unphysical.

Degree-four vertices, those with only four folds emerging from a central vertex, are a special case that has been recently explored \cite{Waitukaitis2019}. The configuration space of a degree-four vertex can be obtained from Fig. \ref{fig:config} by considering a particular planar slice. For example, if we create a degree-four vertex by removing fold $\theta_{02}$ from Fig. \ref{fig:config}a, the configuration space of the degree-four vertex is the intersection of the surfaces in Fig. \ref{fig:config} with the plane $\theta_{02} =0$. This configuration space is, therefore, one dimensional and the two nappes become disconnected for both positive and negative Gaussian curvature.

This reasoning can also be used to explore the configuration spaces of non-triangulated origami. If we are given an arbitrary origami fold pattern, any non-triangular faces can be triangulated, introducing new fold angles, $(\phi_1,\cdots, \phi_M)$. The proper isometries of the non-triangulated origami are then the intersection of the triangulated configuration space with the hyperplane defined by $(\phi_1,\cdots, \phi_M) = 0$. Therefore, the dimension of the configuration space becomes $D = V_b-3-M$, where $M$ is the number of diagonals added to triangulate the fold pattern. Because these hyperplanes pass through the origin (where the origami is unfolded), they do not change the fundamental topology of the configuration spaces of triangulated origami shown in Fig. \ref{fig:config}.

The configuration spaces in Fig. \ref{fig:config}b -- d give us a first picture of the interplay between origami energetics and kinematics. If we imagine that a torsional spring of stiffness $\kappa$ has been placed on each fold of Fig. \ref{fig:config}a, the energy functional would be $E = (\kappa/2) \sum_{i=1}^N \theta_{nN}^2$. The equi-energy surfaces are given by spheres centered on the state with $\theta_{0i}=0$ and so the ground state is the configuration (or configurations) that are closest to the flat state. Appendix \ref{sec:vertices} provides some mathematical machinery to expand this discussion to general origami fold patterns. In addition to determining the kinematics of an origami structure near the flat state, the matrix $M_{i j}^{(n)}$ also determines the fold angles as a function of the vertex heights through
\begin{equation}\label{eq:angle}
\theta_{\sigma(n,i) n} = {\sum_j}^{N(n)} M_{i j}^{(n)} \left(\frac{h_{\sigma(n,j)}-h_n}{L_{\sigma(n,j) n}}\right),
\end{equation}
where $\theta_{\sigma(n,i) n}$ is the fold angle connecting vertex $\sigma(n,i)$ to vertex $n$. Note that the quadratic terms in Eq. (\ref{eq:angle}) actually vanish so this equation is accurate to quadratic order as well.
Using Eq. (\ref{eq:angle}) we can write an energy functional for a nearly-flat origami structure as
\begin{eqnarray}\label{eq:energy}
E &=& \frac{1}{2} \sum_n \sum_{ijk}^{N(n)} \kappa_{n \sigma(n,i)} M_{i j}^{(n)} M_{i k}^{(n)} \\
&&\times \left( \frac{h_{\sigma(n,j)}-h_n}{L_{\sigma(n,j) n}}\right)  \left( \frac{h_{\sigma(n,k)}-h_n}{L_{\sigma(n,k) n}}\right) \nonumber
\end{eqnarray}
where the sum over $n$ is over internal vertices only. Any fold that joins an internal vertex $n$ to a boundary vertex $k$ has torsional stiffness $\kappa_{n k}$ whereas a fold connecting internal vertex $n$ to internal vertex $m$ has stiffness $2 \kappa_{n m}$ because such folds are double counted in Eq. (\ref{eq:energy}).
Thus, for a single vertex with equal fold stiffness $\kappa$ and zero equilibrium fold angles, the decomposition of deformations in terms of collective variables $c_-$ and $c_n$ yields an energy
\begin{equation}
E = \frac{1}{2} \kappa \left[ \lambda_- (c_-)^2  + \sum_n \lambda_n (c_n)^2 \right].
\end{equation}
For $K_0 > 0$, we introduce a new collective variable $\xi$ such that $c_- = \pm K_0 \cosh \xi$ and $c_n = K_0 n_n \sinh \xi$, where $n_n$ are the components of a unit vector. When $K_0 < 0$, we instead use $c_- = |K| \sinh \xi$ and $c_n = \pm |K| n_n \cosh \xi$. Therefore,
\begin{equation}\label{eq:energy1}
E = \frac{\kappa K^2}{2} \left\{ \begin{array}{ccc}
\lambda_- \cosh^2 \xi + \sinh^2 \xi \sum_n \lambda_n (n_n)^2, & & K_0 > 0,\\
\lambda_- \sinh^2 \xi + \cosh^2 \xi \sum_n \lambda_n (n_n)^2, & & K_0 < 0.
\end{array} \right.
\end{equation}
There is an obvious generalization of Eq. (\ref{eq:energy1}) to the case when the fold stiffnesses are not all equal.

For both signs of $K_0$, Eq. (\ref{eq:energy1}) has a minimum at $\xi = 0$. When $K_0 < 0$, this implies $E = \kappa K^2 \lambda_- c_-^2/2$ and is independent of the choice of $n_n$ or the values of $\lambda_n$. There are two energy minima corresponding to the two points closest to the flat state in Fig. \ref{fig:config}c, independent of any other details of the shape.
When $K_0 < 0$, on the other hand, the component of $n_n$ corresponding to the smallest eigenvalue $\lambda_n$ will be $1$ and the remaining components will be $0$. Hence, $E = \kappa K^2 \lambda_{n_{min}} c_{n_{min}}^2/2$, where $n_{min}$ is the index of the smallest eigenvalue.

\section{Conclusions}

To conclude, we have derived the form of the configuration space of non-Euclidean origami for small amounts of Gaussian curvature near the flat state. For single positive Gaussian curvature vertices, the configuration is characterized by nappes that are separated near the flat state, whereas for negative Gaussian curvature, the configuration space remains connected. Though we have analyzed the case of a single degree-$N$ vertex in detail, the procedure we have used can be applied to explore the kinematics and energetics of more complex, nearly flat origami structures with or without Gaussian curvature. We first consider the case of multiple vertices with $K_n > 0$. Around each vertex, Eq. (\ref{eq:mainresult2}) establishes a single equation for $h_n$ as a function of the heights of the vertices surrounding it. We further assume that this equation has two distinct real solutions for $h_n$. Then the analysis in the previous section establishes that no matter how we deform the boundary vertices, there is no way for the configuration of this vertex to pass from one configuration space nappe to the other. The conclusion is that distinct branches of the configuration space of a complex, origami fold pattern that are distinguished by a $K>0$ vertex being on different nappes are topologically disconnected -- if they were not, there would be also be a way of passing from one nappe to the other on a single vertex. Unfortunately, it is difficult to determine whether or not every combination of nappes can be realized when $K > 0$. The case for $K < 0$ is murkier because, while a single vertex remains connected, there is no reason that global constraints might not lead to disconnected components of the configuration space. Indeed, this must be possible in principle, as triangulated fold patterns with disconnected configuration spaces, albeit rare, have been found \cite{silverberg2015origami}.

Finally, we note that this work provides a new mechanism by which the mechanical response of an origami metamaterial sheet can be molded. In principle, an initially flat structure could be stiffened by imposing a small amount of positive Gaussian curvature. Moreover, Gaussian curvature provides a new means of controlling how a responsive origami structure self-folds by separating the individual nappes so that misfolding is significantly less likely. This suggestion will be followed up in a future work.

\begin{acknowledgements}
We acknowledge funding from the National Science Foundation under grant NSF DMR-1822638 and useful conversations with D.W Atkinson, Z. Rocklin and B. Chen. This work was done in part at the Aspen Center for Physics under grant NSF PHY-1607611.
\end{acknowledgements}

\bibliography{noneuclidean}

\newpage

\appendix

\section{Generalizing the formalism to nonzero Gaussian curvature}\label{sec:formalism}
The problem we seek to solve in this paper lies in reconciling the linear and quadratic length-preserving motions. When vertices have Gaussian curvature, the vertices will not typically lie flat. Hence, we would expect them to be well-described by the linear equation Eq. (\ref{eq:linearisom}). As the Gaussian curvature goes to zero, however, quadratic constraints must somehow emerge.

As before, we will approach the analysis of the possible motions by expanding around the flat state. We expect this expansion to be valid so long as the Gaussian curvature of the vertices is sufficiently small. Denoting the planar angles around any vertex with $\alpha_n$, the discrete Gaussian curvature is $K = 2 \pi - \sum_n \alpha_n$. We imagine that the deformation of a structure is governed by an expansion of the form
\begin{equation}
\mathbf{X}_n = \mathbf{X}^{(0)}_n + \mathbf{u}^{(1)}_n + \mathbf{u}^{(2)}_n
\end{equation}
where $\mathbf{X}^{(0)}_n$ is the position of a flattened origami structure and the superscript of $\mathbf{u}$ represents the order in a formal expansion of the displacement.

Because we are expanding the deformations around an otherwise flat structure, the equilibrium lengths of the edge connecting vertex $n$ and $m$ will not be represented by the distances between the planar vertex positions, $\mathbf{X}^{(0)}_n$. Instead, we let $\Delta_{nm} = L_{nm}^2-(\mathbf{X}^{(0)}_{n}-\mathbf{X}^{(0)}_{m})^2$ measure the deviation of the equilibrium edge lengths from the lengths of the edges when projected to the $xy-$plane. We denote $\bm{\Delta}$ the vector formed by concatenating the components $\Delta_{nm}$ for each edge. We similarly write $\mathbf{u}^{(1)}$ and $\mathbf{u}^{(2)}$ as the concatenation of the vertex displacements at first and second order. Finally, introduce a quadratic function, $\mathbf{f}(\mathbf{u})$ with components $(\mathbf{u}_n - \mathbf{u}_m)^2$ for each edge, $(n,m)$. Then we have
\begin{equation}\label{eq:fullexpansion}
\bm{\Delta} = \mathbf{R} \mathbf{u}^{(1)} + \mathbf{R} \mathbf{u}^{(2)} + \mathbf{f}(\mathbf{u}^{(1)}),
\end{equation}
where $\mathbf{R}$ is the compatibility matrix mapping vertex displacements to linear changes in the edge lengths \cite{connelly1980rigidity,lubensky2015phonons}.

To linear order, one should solve $\bm{\Delta} = \mathbf{R} \mathbf{u}^{(1)}$. However, this linear equation can only have a solution if the left-hand side of the equation lies in the image of $\mathbf{R}$. We denote the projection of a vector into the image of $\mathbf{R}$ with a subscript $||$, and a projection into the orthogonal complement $\perp$. Therefore, Eq. (\ref{eq:fullexpansion}) decomposes into the pair
\begin{eqnarray}\label{eq:linearEq}
\bm{\Delta}_{||} &=& \mathbf{R} \mathbf{u}^{(1)} + \mathbf{R} \mathbf{u}^{(2)} + \mathbf{f}_{||}(\mathbf{u}^{(1)})\\
\bm{\Delta}_{\perp} &=& \mathbf{f}_{\perp}(\mathbf{u}^{(1)}) \label{eq:sigmaEq}
\end{eqnarray}

Eq. (\ref{eq:linearEq}) can now be solved order by order. To first order, $\mathbf{u}^{(1)} = \mathbf{u}_{||} + \mathbf{h}$, where $\mathbf{u}_{||}$ is any solution of $\mathbf{R} \mathbf{u}_{||} = \bm{\Delta}_{||}$, and $\mathbf{h}$ is in the right null space of $\mathbf{R}$. At the next order, we obtain a correction $\mathbf{R} \mathbf{u}^{(2)} = - \mathbf{f}_{||}(\mathbf{u}_{||} + \mathbf{h})$.

Since we are expanding around a flat origami structure, we can further restrict the structure of $\mathbf{u}_{||}$ and $\mathbf{h}$. Particularly, it must be that $\mathbf{h}$ can only involve the three in-plane Euclidean motions and the vertical displacements of all of the vertices. Consequently, $\mathbf{u}_{||}$ can be chosen so that the vertex displacements lie in the $xy-$plane and $\mathbf{h}$ can then contain only vertex displacements along the $\hat{\mathbf{z}}$.

Eq. (\ref{eq:sigmaEq}) is not dispensed with so easily. It remains a quadratic constraint on $\mathbf{h}$ of the form
\begin{equation}\label{eq:quadraticEq}
\bm{\Delta}_{\perp} = \mathbf{f}_{\perp}(\mathbf{u}_{||}+\mathbf{h}) =  \mathbf{f}_{\perp}(\mathbf{u}_{||}) +  \mathbf{f}_{\perp}(\mathbf{h}).
\end{equation}
The last equality follows from the fact that $\mathbf{u}_{||}$ is perpendicular to $\mathbf{h}$ and $\mathbf{f}$ is quadratic. Finally, we neglect $\mathbf{f}_\perp(\mathbf{u}_{||})$ since it is quadratic in $|\bm{\Delta}_{||}|$. This is valid when $|\bm{\Delta}_{\perp}| \sim |\bm{\Delta}_{||}|$.

To interpret Eq. (\ref{eq:quadraticEq}), we let $\{\bm{\sigma}_1, \bm{\sigma}_2, \cdots\}$ be the basis of wheel stresses of $\textrm{ker}~\mathbf{R}^T$ described in Ref. \cite{chen2018}. In this basis,
\begin{equation}
\bm{\sigma}_n \cdot \mathbf{f}(\mathbf{h}) \approx \bm{\sigma}_n \cdot \bm{\Delta},
\end{equation}
where the left-hand side can be interpreted as the discrete Gaussian curvature at vertex $n$, or alternatively as a quadratic form, $\mathbf{h}^T \mathbf{Q}_n \mathbf{h}$ \cite{chen2018}. Finally,
\begin{equation}\label{eq:gcEQ1}
\mathbf{h}^T \mathbf{Q}_n \mathbf{h} = \bm{\sigma}_n \cdot \bm{\Delta} \equiv K_n.
\end{equation}
We note that, when $K_n=0$, Eq. (\ref{eq:gcEQ1}) reproduces the results of Chen \textit{et al.} \cite{chen2018} for flat origami.
Notice that the right-hand side of Eq. (\ref{eq:gcEQ1}) involves only lengths of the bonds, encoded through $\bm{\Delta}$. This is then a discrete version of Gauss' \textit{theorema egregium}, which relates the Ricci curvature on a surface -- a completely intrinsic quantity -- to the Gaussian curvature -- an extrinsic quantity.

\subsection{Relation to linear analysis}
Rather than expanding the deformations around a nearly flat state. we could have solved Eq. (\ref{eq:linearisom}) directly from a slightly deformed state. Here, we demonstrate that our approach yields the same results to linear order. Let $\mathbf{X}_n = \mathbf{X}_n^{(0)} + h_n \hat{\mathbf{z}}$, where $\mathbf{X}_n^{(0)}$ has no $\hat{\mathbf{z}}$ component. Similarly, write $\mathbf{u}_n = \mathbf{w}_n + h_n^{(1)} \hat{\mathbf{z}}$, where $\mathbf{w}_n$ has no $\hat{\mathbf{z}}$ component. Eq. (\ref{eq:linearisom}) then reads
\begin{equation}
2 \left(\mathbf{X}^{(0}_n - \mathbf{X}^{(0)}_m\right) \cdot \left(\mathbf{w}_n - \mathbf{w}_m\right) + 2 ({h}_n^{(0)} - {h}_m^{(0)})  ({h}_n^{(1)} - {h}_m^{(1)}) = 0.
\end{equation}
Let $\sigma_i^{nm}$ be a wheel stress around vertex $i$. Then we have
\begin{equation}
\sum_{nm} \sigma_i^{nm} 2 ({h}_n^{(0)} - {h}_m^{(0)}) ({h}_n^{(1)} - {h}_m^{(1)}) = 0,
\end{equation}
where the sum is over all edges. Rewriting this in terms of the concatenated vectors $\mathbf{h}$, we obtain
\begin{equation}\label{eq:gcEQA}
\left(\mathbf{h}^{(0)}\right)^T \mathbf{Q}_i \mathbf{h}^{(1)} = 0.
\end{equation}

Alternatively, if we expand Eq. (\ref{eq:gcEQ1}) around $\mathbf{h}^{(0)}$ which satisfies ${\mathbf{h}^{(0)}}^T \mathbf{Q}_i \mathbf{h}^{(0)} = K_i$, we also obtain Eq. (\ref{eq:gcEQA}).

\section{Single vertices}\label{sec:vertices}

\begin{figure}
\includegraphics[width=3.5in]{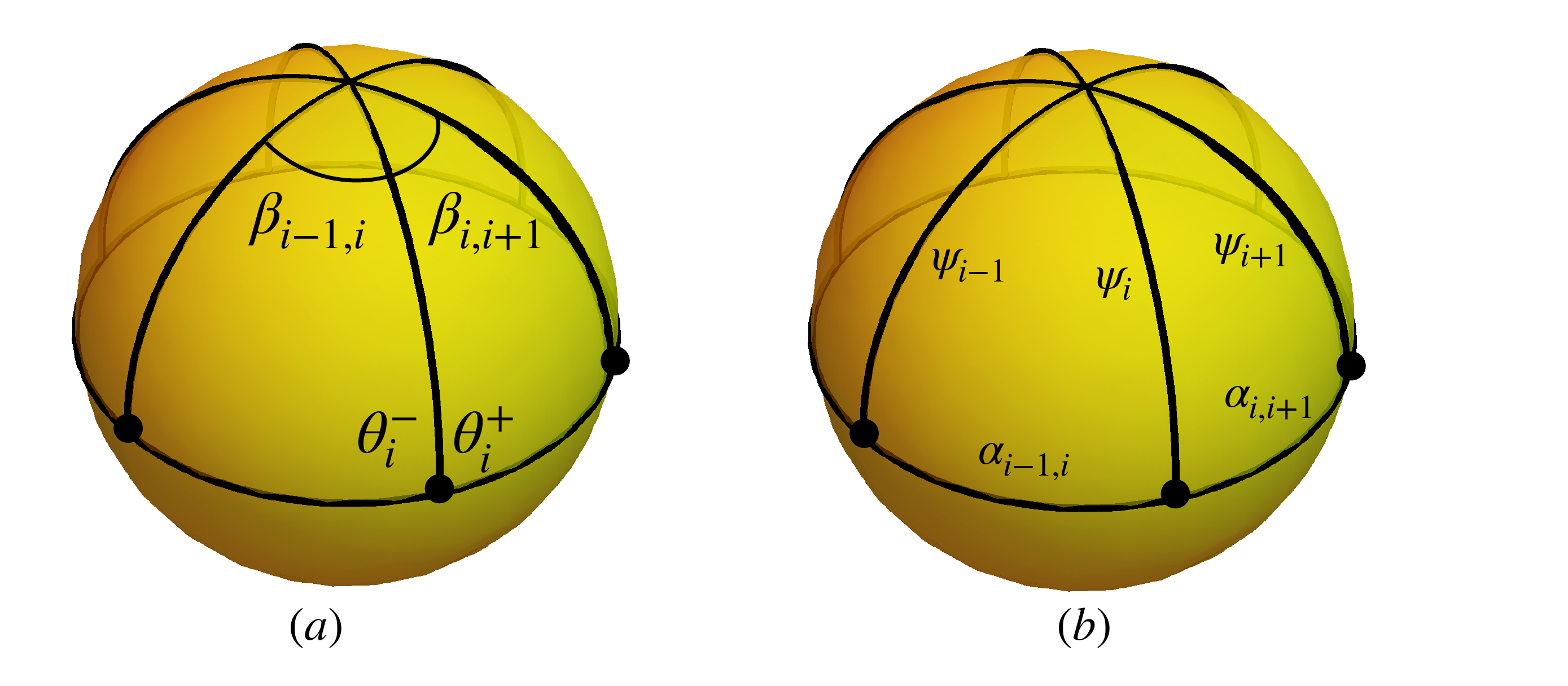}
\caption{\label{fig:sphericaltrig} The intersection of a sphere with a vertex at its center is a spherical polygon, which we decompose into triangular slices as shown. (a) The dihedral angle of the $i^{th}$ fold is $\theta_i^+ + \theta_i^-$. (b) The side lengths are the planar angles $\alpha_{i,i+1}$ and the angle the folds make with respect to the $xy-$plane, $\psi_i$.}
\end{figure}

We denote the central vertex with $0$ and number the boundary vertices from $n=1$ to $N$. Denote $\alpha_{n,n+1}$ as the angle between fold $n$ and $n+1$, interpreted assuming $\alpha_{N,N+1} = \alpha_{N,1}$, and assume that  $\alpha_{n,n+1}$ is always between $0$ and $\pi$.
Since we are interested in single vertices near the flat state, it is useful to change variables from the vertex heights to the angles made by the folds with respect to the $\hat{\mathbf{z}}$ axis, oriented with respect to the reference $z-$axis: $\psi_n = \pi/2 + (h_0 - h_n)/L_{n 0}$ where $L_{n0}$ is the length of fold $n$.

It is well known that a single vertex can be interpreted as a spherical polygon in which the side lengths are given by the planar angles $\alpha_{n,n+1}$ and the dihedral angles by the interior angles of the polygon (Fig. \ref{fig:sphericaltrig}); this connection has been used to explore the full configuration space of single origami vertices in general \cite{kapovich1997hodge,streinu2004}. Fig. \ref{fig:sphericaltrig} shows that such a polygon can be decomposed into triangular slices. Spherical trigonometry then allows one to write the dihedral angles entirely in terms of the $\psi_n$. For small deformations, these $N$ angles $\psi_n = \pi/2 + \delta \psi_n$ where $\delta \psi_n = (h_n - h_0)/L_{n0}$.
Finally, we define $\theta_n$ as the dihedral angle made by the $n^{th}$ fold; the diagram in Fig. \ref{fig:notation} shows that $\theta_n = \theta_n^+ + \theta_n^-$. Finally, we let $\theta_n = \pi - \delta \theta_n$ and assume $\delta \theta_n$ is small.

Expanding to quadratic order, we obtain the linear relationship
\begin{equation}
\delta \theta_n = \sum_m {M}_{n m} \delta \psi_m
\end{equation}
where ${M}_{nm} = -\csc \alpha_{n,n+1} \delta_{n,m+1} - \csc \alpha_{n-1,n} \delta_{n,m-1} + (\cot \alpha_{n,n+1} + \cot \alpha_{n-1,n}) \delta_{nm}$. 
Expanding the angles $\beta_{n,n+1}$ around $\alpha_{n,n+1}$ and using $\sum_n \beta_{n,n+1} = 2\pi$, we also find an expression for the Gaussian curvature of the vertex, $K = 2 \pi - \sum_n \alpha_{n,n+1}$,
\begin{equation}\label{eq:gcvertex}
K = -\frac{1}{2} \sum_{nm} \delta \psi_n \delta \psi_m {M}_{nm}.
\end{equation}
Comparing Eq. (\ref{eq:gcvertex}) to Eq. (\ref{eq:mainresult}) provides a connection between the matrix $\mathbf{Q}$ governing the configuration space in terms of the vertex heights to the matrix $\mathbf{M}$, having components $M_{nm}$, governing the configuration space in terms of angles $\delta \psi_n$. In particular, while $\mathbf{Q}$ should have an additional zero eigenvalue from global translations of the vertex in the $\hat{\mathbf{z}}$ direction, it shares the same number of positive and negative eigenvalues as $\mathbf{M}$ \cite{chen2018}.

%
%

\end{document}